\documentclass[aps,prl,superscriptaddress,showpacs,twocolumn]{revtex4}
\usepackage{amsfonts}
\usepackage{amsmath}
\usepackage{amssymb}
\usepackage{graphicx}

\begin{document}

\title{Time reversal Aharonov-Casher effect in mesoscopic rings with Rashba spin-orbital interaction}
\author{Zhenyue Zhu}
\affiliation{Department of Physics, Oklahoma State University, Stillwater, OK USA}
\author{Yong Wang}
\affiliation{Institute of Physics, Chinese Academy of Sciences, Beijing 100080, China}
\author{Ke Xia}
\affiliation{Institute of Physics, Chinese Academy of Sciences, Beijing 100080, China}
\author{X. C. Xie}
\affiliation{Department of Physics, Oklahoma State University,
Stillwater, OK USA}
\affiliation{Institute of Physics, Chinese
Academy of Sciences, Beijing 100080, China}
\author{Zhongshui Ma}
\affiliation{School of Physics, Peking University, Beijing 100871, China}

\begin{abstract}
The time reversal Aharonov-Casher (AC) interference effect in the
mesoscopic ring structures, based on the experiment in Phys. Rev.
Lett. \textbf{97}, 196803 (2006), is studied theoretically. The
transmission curves are calculated from the scattering matrix
formalism, and the time reversal AC interference frequency is
singled out from the Fourier spectra in numerical simulations. This
frequency is in good agreement with analytical result. It is also
shown that in the absent of magnetic field, the
Altshuler-Aronov-Spivak type (time reversal) AC interference retains
under the influence of strong disorder, while the Aharonov-Bohm type
AC interference is suppressed.
\end{abstract}

\date{\today}
\pacs{73.23.-b, 71.70.Ej, 72.15.Rn} \maketitle

\medskip

The advancement of spintronics and spintronic materials has provided
opportunities to study and utilize new electronic devices based on
the electron spin degrees of freedom\cite{spin}. The novel spin
related properties can be detected by tuning the spin-orbital
interaction (SOI). Phenomena related to Rashba effects in, e.g.,
InAs and In$_{1-x}$Ga$_{x}$As based two-dimensional electron gas
(2DEG) systems have been observed\cite{sato,nitta}. The value of the
Rashba parameter in these semiconductor systems can be as high as
$4\times 10^{-11}$eVm. The experimental results\cite{sato,nitta}
have shown that in InAs and In$_{1-x}$Ga$_{x}$As based 2DEG systems,
the Rashba effect is responsible for the spontaneous spin splitting.
Many advanced electronic devices have been proposed to manipulate
electron spin, such as spin transistor\cite{datta},
waveguides\cite{wang,sun}, spin filters\cite{koga}, spin
interferometers\cite{Nitta1,shen}, etc. In the spin interference
device\cite{Nitta1,shen,frus}, the resistance is adjusted
\textit{via} the gate voltage using the spin interference phenomenon
of the electronic wave functions in a mesoscopic ring. In the
presence of Rashba SOI\cite{Rashba}, the wave function acquires the
Aharonov-Casher (AC) phase when an electron moves along the
ring\cite{AC1,AC2}. If the quantum coherence length is larger than
or comparable with the electron mean free path, the resistance
amplitude will depend on the AC phases of electron wave functions in
the two arms of the ring and display interference pattern.
Meanwhile, the modulation of interference with the aid of AC phase
could be further tuned through an applied gate voltage\cite{nitta}.

The experimental demonstrations of the spin interference in the AC
rings have been reported recently\cite{Konig,bergsten}. As expected,
the resistance of the ring oscillates with the gate voltage and an
external magnetic field. However, their experimental results do not
directly establish the time reversal AC frequency. In this study, we
numerically investigate the transport properties of multi mesoscopic
rings with magnetic field and Rashba SOI in the framework of
Landauer-B\"{u}ttiker formalism. The interference modes are analyzed
by means of the fast Fourier transformation (FFT) algorithm. As a
subsidiary interference, the magnetic flux leads to the conventional
Aharonov-Bohm (AB) \cite{AB1} and Altshuler-Aronov-Spivak (AAS)
\cite{AAS1,AAS2} effects. Subtracting the contribution of magnetic
flux and determining the pathes of electrons from the frequency
spectra, AB and AAS type oscillations associated with SOI in rings
are confirmed. In the presence of dephasing, AAS type (time
reversal) AC interference remains, while the AB type oscillation is
suppressed.

\begin{figure}[th]
\begin{center}
\includegraphics[width = 6.0cm, angle=0, bb = 0 0 300 255]{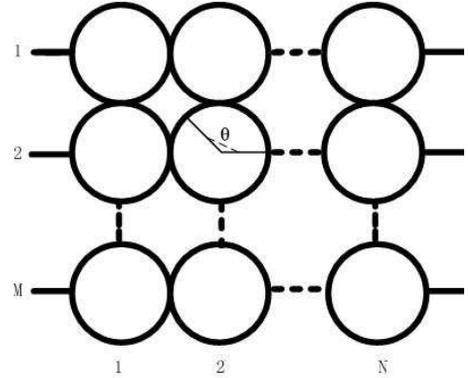}
\end{center}
\caption{The schematic graph of the M-N ring-array structure. The
dot lines represent the omitted rings. $\protect\theta $ is the
local polar coordinates.} \label{M-N}
\end{figure}

The M-N ring-array structure, as shown in Fig. \ref%
{M-N}, mimics the experimental configurations \cite{bergsten}. For
simplicity, each arm of the ring is assumed to be a one-dimensional
(1D) quantum wire and connected with each other at crossing points.
The $2M$ leads are connected to the M-N ring-array as $M$
1D-channels. The external magnetic and electric fields are
perpendicular to the plane of the rings. The system can be solved by
dividing it into several sections, i.e., the semi-infinite left
M-leads, the region of M-N rings array in the presence of SOI, and
the semi-infinite right M-leads. With the continuity condition on
the boundaries between sections, the scattering matrix
formalism\cite{andreas} can be established.

Because of the absence of the SOI in the leads, the wavefunction in the left
(right) i-th channel can be written in the form of $\psi
_{iL}=a_{iL,+}e^{ikx}|\uparrow >+a_{iL,-}e^{ikx}|\downarrow
>+b_{iL,+}e^{-ikx}|\uparrow >+b_{iL,-}e^{-ikx}|\downarrow >$ and $\psi
_{iR}=a_{iR,+}e^{ikx}|\uparrow >+a_{iR,-}e^{ikx}|\downarrow >$ for an
incident electron with momentum $\hbar k$. The coefficients satisfy the
condition $\sum_{i\sigma =\pm }\left\vert b_{iL,\sigma }\right\vert
^{2}+\sum_{i\sigma =\pm }\left\vert a_{iR,\sigma }\right\vert
^{2}=\sum_{i\sigma =\pm }\left\vert a_{iL,\sigma }\right\vert ^{2}=1$.

In the region of M-N rings array, the electronic wavefunction $\Psi
$ in the rings can be expanded in terms of the complete set of
eigenfunctions $\psi _{n\sigma }\left( \theta \right) $ as $\Psi
_{ij}\left( \theta ;k\right) =\sum_{n\sigma }c_{n\sigma }\psi
_{n\sigma }\left( \theta ;k\right) $ with the spin index $\sigma$
for the ring (i, j). For a single ring, the Hamiltonian in the local
polar coordinates reads\cite{Hamiltonian}
\begin{eqnarray}
H &=&\frac{\hbar ^{2}}{2m^{\ast }r^{2}}\left( -i\partial _{\theta
}+\frac{\Phi }{\Phi _{0}}\right) ^{2}+\frac{\alpha }{2r}\left(
\begin{array}{cc}
0 & -e^{-i\theta } \\
e^{i\theta } & 0%
\end{array}%
\right)  \notag \\
&&+\frac{\alpha }{r}\left(
\begin{array}{cc}
0 & e^{-i\theta } \\
e^{i\theta } & 0%
\end{array}%
\right) \left( -i\partial _{\theta }+\frac{\Phi }{\Phi _{0}}\right)
\label{Hamilton}
\end{eqnarray}%
(Here $m^{\ast }$ is the electron effective mass and we take
$m^{\ast }=0.05m $, $r$ is the radius of a single ring, $\alpha $ is
the Rashba SOI strength, and $\Phi $ is the magnetic flux in unit of
the flux quanta $\Phi _{0}=h/e$. We ignore the Zeeman terms because
of the weak magnetic field, and concentrate on the effect generated
by SOI.) The eigenfunctions of the system are $\psi _{n,+}\left(
\theta \right) =\left( 2\pi \right) ^{-1/2}e^{in\theta }\left( \cos
\beta /2,-e^{i\theta }\sin \beta /2\right) ^{T}$, $\psi _{n,-}\left(
\theta \right) =\left( 2\pi \right) ^{-1/2}e^{in\theta }\left( \sin
\beta /2,e^{i\theta }\cos \beta /2\right) ^{T}$, and the
corresponding eigenvalues are $\epsilon _{n,\pm }=\left( \hbar
^{2}/2m^{\ast }r^{2}\right) \left[ \left( n+\Phi /\Phi _{0}\right)
^{2}+\left( 2n+2\Phi /\Phi _{0}+1\right) \Omega _{\pm }\right] \mp
\left( \alpha /2r\right) \left( 2n+2\Phi /\Phi _{0}+1\right) \sin
\beta $, with $\tan \beta =2mr\alpha /\hbar ^{2}$, $\Omega _{\pm
}=-1/2\left( 1\pm \cos \beta \right) $, and $T$ denotes the
transposition of vectors.

The linear equations among coefficients $c_{n\sigma }$ can be
determined by the Griffith conditions \cite{Griffith} at each joint
point. After obtaining the sets of wavefunctions $\psi _{iL}$, $\psi
_{iR}$ and $\Psi _{ij}\left( \theta ;k\right) $, the scattering
matrices can be derived according to the multi-mode
scattering-matrix procedure given in Ref. \cite{andreas}. Therefore,
the total transmission and reflection coefficients of ring-array
system can be obtained.

\begin{figure}[tbp]
\begin{center}
\includegraphics[height=2.200in, width=3.200in]{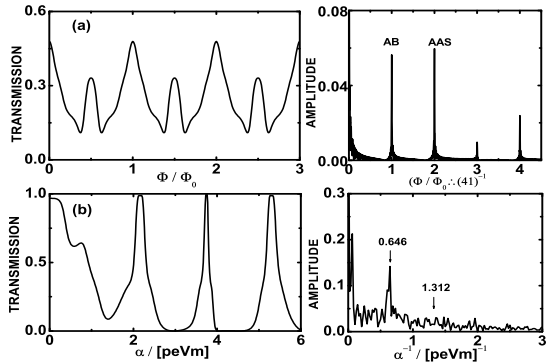}
\end{center}
\caption{(a) The curves for T vs $\Phi$ and the corresponding
Fourier spectrum for 3-3 ring-array with $k=1.009651\times 10^{8}m^{-1}$ and $%
\protect\alpha = 2peVm$; (b) The curves for T vs $\protect\alpha$
and the corresponding Fourier spectrum for 3-3 ring-array with
$k=1.009651\times 10^{8}m^{-1}$ and $\Phi = 0$.} \label{3-3}
\end{figure}

We have calculated the transmission coefficients for several
ring-array systems under different magnetic flux $\Phi $ and Rashba
coefficient $\alpha $. The phenomena are similar in the systems with
different number of rings and radii, and we take the 3-3 ring-array
with $r=1\mu $m as a special case to display our results. In Fig.
$\ref{3-3}$(a), we present the transmission coefficient versus
magnetic flux ($T$ vs $\Phi $) of the 3-3 ring-array with the
electronic incident wavevector $k=1.009651\times 10^{8}$m$^{-1}$ and
the Rashba coefficient $\alpha =2$peVm. The transmission coefficient
oscillates with the magnetic flux periodically. Although the shapes
of the curves may look quite different with different $k$ and
$\alpha $, the periods of the curves are the same. In fact, as we
can see in the right panel, the FFT spectra of the curve reveals
different interference modes (namely AB, AAS, \textit{etc.}). For
various systems with different $k$ and $\alpha $, the FFT spectra of
the $T$ vs $\Phi$ curves include the same modes, but the amplitude
of these modes will be different.

In the absence of magnetic flux, transmission $T$ oscillates with
tuning the SOI strength [Fig. $\ref{3-3}$(b)]. The oscillation is no
longer strictly periodic at large incident electron momentum,
because of the complicated back and forth multi-scattering between
channels and rings. However, the Fourier spectra still reveal the
same main frequency at $\alpha ^{-1}=0.646$ $\left(
\text{peVm}\right) ^{-1}$, which corresponds to the AB type
oscillation. Besides this mode, other modes at $\alpha ^{-1}=1.312$
$\left( \text{peVm}\right)^{-1}$,  $...$ (AAS type \textit{etc.}),
are not very clear here. While in the case for a single AC ring,
these modes will be clearly shown in the FFT spectra.

The frequencies subsumed in the Fourier spectra on the right panels
of Fig. $\ref{3-3}$(b) can be understood analytically from a single
ring structure. The AC phase difference of opposite spins traveling
in full counter-clockwise (CCW) and clockwise (CW) circle are given
by \cite{frus} $\Delta \varphi _{\psi _{-}^{\downarrow }-\psi
_{+}^{\uparrow
}}=2\pi \left[ 1+\sqrt{1+\left( 2rm^{\ast }\alpha /\hbar ^{2}\right) ^{2}}%
\right] $ and $\Delta \varphi _{\psi _{-}^{\uparrow }-\psi _{+}^{\downarrow
}}=2\pi \left[ 1-\sqrt{1+\left( 2rm\alpha /\hbar ^{2}\right) ^{2}}\right] $.
In the limit of large $\alpha $, the two phase differences take the same
form as in Ref.\cite{Nitta1,Konig} . Therefore, the AAS oscillation
amplitude can be expressed as \cite{Nitta3}
\begin{equation}
\delta R_{\alpha }^{AAS}=\delta R_{\alpha =0}\cos \left[ 2\pi \sqrt{1+\left(
\frac{2rm^{\ast }\alpha }{\hbar ^{2}}\right) ^{2}}\right] .  \label{AAS}
\end{equation}%
From this relation the AAS type (time reversal) AC frequency is
approximately at $\alpha _{AAS}^{-1}=2rm^{\ast }/\hbar
^{2}=1.31$peVm, while the AB type oscillation frequency is $\alpha
_{AB}^{-1}=rm^{\ast }/\hbar ^{2}=0.65$peVm, as indicated in the
Fourier spectrum.

In order to emphasize on the AAS effect in the ring structure and
bring light to the objective law governing the spin interference, we
introduce disorder induced dephasing effect, so that the AB type
oscillation is suppressed. For simplicity, a single ring with two
leads is considered, which is enough to manifest the main principle
of AAS effect in a ring structure. The conductance is numerically
calculated with a tight-binding model. The tight-binding Hamiltonian
for a ring in Eq. (1) is given by\cite{nikolic}
\begin{eqnarray*}
H &=&\sum_{l=1}^{N}\sum_{\sigma =\uparrow \downarrow }\epsilon
_{l}c_{l\sigma }^{\dag }c_{l\sigma }-\sum_{l=1}^{N}\sum_{\sigma =\uparrow
\downarrow }(tc_{l\sigma }^{\dag }c_{l+1\sigma }+h.c.) \\
&&-\sum_{l=1}^{N}[i\lambda (e^{-i\varphi _{l}}c_{l,\uparrow }^{\dag
}c_{l+1,\downarrow }+e^{i\varphi _{l}}c_{l,\downarrow }^{\dag
}c_{l+1,\uparrow })+h.c.],
\end{eqnarray*}%
where $\epsilon _{l}$ is the lattice on site energy, the operator $%
c_{l,\sigma }^{\dag }(c_{l,\sigma })$ creates (annihilates) an electron with
spin $\sigma $ at site $l$, $N$ is the total number of lattices in the ring,
and $\varphi _{l}=\pi (2l-1)/N$ as defined in Ref. \cite{nikolic}. The
hopping coefficients in the second and third summations can be expressed as $%
t$ (and $\lambda $) $=t_{0}$ (and $\lambda _{0}$) $\exp (ie\Phi /\hbar N)$
and $t_{0}\equiv \hbar ^{2}/2m^{\ast }a^{2}$ is the nearest neighbor hopping
element with lattice spacing $a$ along the 1D ring. $\lambda _{0}=\alpha /2a$
describes the strength of the Rashba SO interaction. $\Phi $ is the magnetic
flux through the 1D ring. In our numerical calculations, $\epsilon _{l}=0$
if disorder potential at each site is absent, otherwise, the lattice on site energy $%
\epsilon _{l}$ is generated by a uniform distribution $%
[-w/2,w/2]$ of disorder strength $w$. Because of the periodic boundary
condition, $c_{N+1,\sigma }^{\dag }$ ($c_{N+1,\sigma }$) is identical as $%
c_{1,\sigma }^{\dag }(c_{1,\sigma })$. In addition, two semi-infinite 1D
leads are attached to the 1D ring to make sure that the upper and lower arms
have the same length. The influence of these two leads is taken into account
through self-energy terms. The conductance is calculated numerically as
outlined in Ref. \cite{AC4} .

For a certain Fermi energy and Rashba SOI strength, the conductance
of 3-3 rings array exhibits the periodic oscillation as a function
of the magnetic flux $\Phi $. FFT analysis displays AB, AAS, and
other higher frequencies as shown in Fig. $\ref{3-3}$(a). However,
as we will show below, with the increasing of the disorder strength
$w$, the AB oscillation is suppressed and only the time reversal AAS
oscillation amplitude remains. Similar results have been reached in
a 1D square loop system \cite{AC4}.

To see the suppression of AB type oscillation in the presence of
disorder, the conductance is calculated for a single 1D ring with
different Rashba SOI strengths. Employing FFT and inverse FFT
techniques, we first calculate the flux dependent conductance and
then extract the AAS oscillation amplitudes at zero magnetic field
for a series of different $\lambda _{0}$. Afterwards we retrieve the
information about AAS oscillation amplitudes vs $\lambda _{0}$ (same
steps as in Ref. \cite{bergsten}). The purpose of this approach is
to single out the time reversal AC phase differences. Under the
influence of magnetic field and Rashba SOI, two paths will
accumulate both AB and AC phases before interference. Since AAS
oscillation is caused by the interference of two time reversal paths
(one CCW circle and another CW circle), it is feasible to find out
AC phase difference between these two time reversal paths, if we
extract AAS oscillation and set magnetic field to be zero.

\begin{figure}[th]
\begin{center}
\includegraphics[width=8.5cm,angle=0]{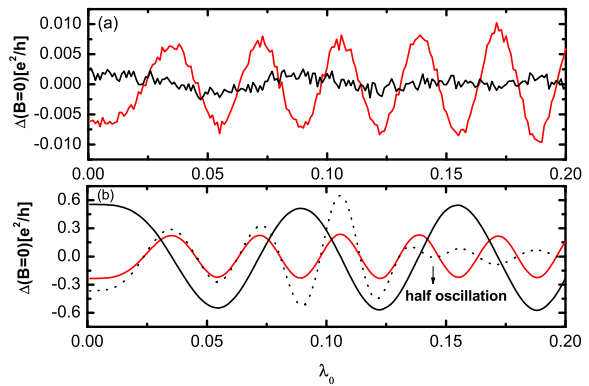}
\end{center}
\caption{ The AAS and AB oscillation amplitudes versus $\lambda_0$
curves are drawn in red and black solid line respectively, for a
single 1D ring (a) under disorder average with $w=0.8$, $E_{f}=-1.8$
and (b) under Fermi energy average. The AAS oscillation amplitudes
versus $\lambda_0$  at $E=-1.8$ (black dotted line) has also been
plotted for comparisons in (b). Other parameters are set as $N=100$,
$t_0=1$. All the energies are measured in the units $t_0$.}
\label{AAS-alpha}
\end{figure}

Fig. $\ref{AAS-alpha}$(a) shows the AAS oscillation amplitude versus
$\lambda_{0}$ curves (red solid lines) and the AB oscillation
amplitude versus $\lambda _{0}$ curves (black solid lines) under
disorder average with disorder strength $w=0.8$. Compared with the
AAS oscillation, the AB oscillation has been suppressed greatly
under the disorder average.

While Fig. $\ref{AAS-alpha}$(b) shows the AAS oscillation amplitude
versus $\lambda_0 $ curves (red solid lines) and the AB oscillation
amplitude versus $\lambda _{0}$ curves (black solid lines) under
average over incident electron energy (Fermi energy). In this case,
the AB oscillation has not been suppressed. Besides, although these
two kinds of averages lead to the similar AAS oscillation shapes in
the conductance, the oscillation amplitude under the disorder
average is almost 20 times smaller than that of energy average case.
The suppression of the amplitude is due to the fact that electron
will be localized at large disorder strength.

In these plots, no trace of half oscillation, which appears in the
experimental results\cite{bergsten}, could be found. However, from
the AAS oscillation amplitudes versus $\lambda _{0}$ for a single 1D
ring at $E=-1.8$ [dotted line in Fig. $\ref{AAS-alpha}$(b)], we
notice that the so-called "half oscillation" appears indeed as the
experiments\cite{bergsten}. In comparison of the AAS oscillation
amplitude versus $\lambda _{0}$ for $E=-1.8 $ and under Fermi energy
average, it is shown that at the strength of SOI where the half
oscillation happened, the following oscillation amplitude differs a
phase shift $\pi $ with the energy averaged oscillation amplitude.
When $\lambda _{0}$ is smaller than 0.145 (before half oscillation),
the peak and dip features of these two curves match well with each
other. But when $\lambda _{0}$ is larger than 0.145, the peak
feature of one curve corresponds to the dip feature of another.
These conclusions can also be inferred from the
experiment\cite{bergsten}. Therefore, we conclude that half
oscillation only happens at some certain Fermi energy (carrier
density). But this behavior disappears after ensemble average
(disorder or Fermi energy average).
\begin{figure}[tbp]
\begin{center}
\includegraphics[height=3.400in, width=3.200in]{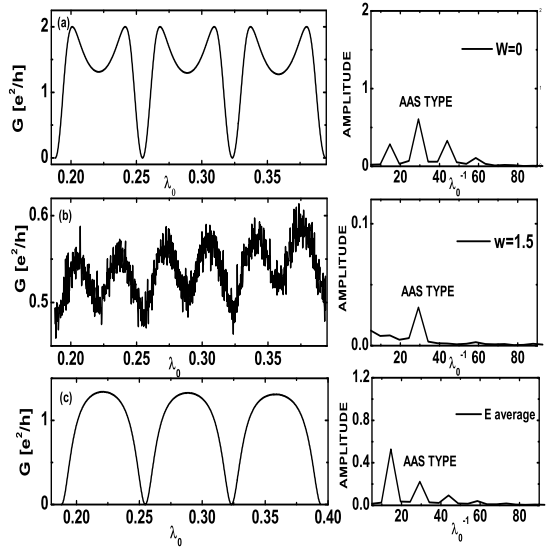}
\end{center}
\caption{The conductance of a single 1D ring versus
$\protect\lambda_0 $ without magnetic field. (a), (b) are calculated
from $E_{f}=-0.05$ with disorder strength $w=0,1.5$ respectively.
(c) is calculated from Fermi energy average without disorder. All
the corresponding FFT spectrum are shown on the right side. The size
of the ring N=100. The theoretical time reversal (AAS type) AC
frequency is at $\lambda_0^{-1}=31.83$. In our FFT spectrum, the AAS
type AC frequency is located at $\lambda_0^{-1}=29.30$.} \label{4}
\end{figure}

From the FFT spectrum of AAS oscillation amplitude versus
$\lambda_0$ curves [red curves in Fig. 3(a) and 3(b)], we find a
frequency peak located at $\lambda_0^{-1}=31.25$. From the
theoretical time reversal AC frequency $\alpha^{-1}=2rm^*/\hbar^2$,
we get $\lambda_0^{-1}=N/\pi$. Thus the theoretical time reversal AC
frequency should be $\lambda_0^{-1}=31.83$ for N=100. Comparing
these two values, our numerical result agrees well with the
theoretical value.

Finally, in order to substantiate our conclusion, we plot the
conductance versus $\lambda _{0}$ in absent of magnetic field for
several cases: without disorder, with disorder average and with
energy average (Fig. $\ref{4}$). From their FFT spectra, it is seen
that different oscillation frequencies are included when disorder is
absent [Fig. \ref{4}(a)]. With strong disorder strength, the AB type
component is suppressed [Fig. 4(b)]. Thus, we are sure that the time
reversal AC effect have been successfully demonstrated in such a
system. However, the case of Fermi energy average is quite different
from disorder average. In the FFT spectrum of energy average [Fig.
\ref{4}(c)], other oscillation frequencies (two half circle
interference ...) can not be smeared out under energy average. The
AB oscillation amplitude at zero magnetic field becomes smaller with
increasing disorder strength, but it still remains strong under
Fermi energy average, which supports our conclusion in Fig.
$\ref{AAS-alpha}$.

In summary, we have numerically demonstrated the time reversal AC
interference effect in mesoscopic rings. In the absent of disorder,
various AC interference patterns are included in the rings.
Nevertheless, most interference modes are weak against disorder,
except the time reversal mode.

\begin{acknowledgments}
This work is supported in part by (ZZ and XCX) DE-FG02-04ER46124
and NSF CCF-0524673; (YW and KX) NBRPC2006CB933000 and NNSFC
10634070; (ZSM) NNSFC 10674004 and NBRPC 2006CB921803.
\end{acknowledgments}

\end{document}